\documentclass[3p,times,procedia]{elsarticle}
\usepackage{nupha_ecrc}

\volume{00}

\firstpage{1}

\journalname{Nuclear Physics A}

\runauth{Vipul Bairathi}

\jid{nupha}

\jnltitlelogo{Nuclear Physics A}

\usepackage{amssymb}

\usepackage[figuresright]{rotating}

\begin{document}

\begin{frontmatter}



\dochead{}

\title{Study of the bulk properties of the system formed in Au+Au collisions at $\sqrt{s_{NN}}$ = 14.5 GeV using the STAR detector at RHIC}


\author{Vipul Bairathi (For the STAR collaboration)$^{1}$}
\footnote{A list of members of the STAR Collaboration and acknowledgments can be found at the end of this issue.}

\address{National Institute of Science Education and Research, Jatni, Orissa, India - 752050}

\begin{abstract}
We present the first measurements of the transverse momentum spectra and azimuthal anisotropy of the identified hadrons ($\pi^{+}$, $\pi^{-}$, $K^{+}$, $K^{-}$, $p(\bar{p})$, $K^{0}_{s}$, $\Lambda(\bar{\Lambda})$ and $\phi$) at mid-rapidity in Au+Au collisions at $\sqrt{s_{NN}}$ = 14.5 GeV for various collision centralities. These measurements are compared to corresponding results from other BES energies. The bulk properties of the system, like the chemical and kinetic freeze-out conditions and the collectivity extracted from the measured yields of the produced particles are presented. The difference between baryon and anti-baryon elliptic flow for minimum bias collisions previously reported by STAR is also observed in the new data taken at $\sqrt{s_{NN}}$ = 14.5 GeV. Furthermore, the new data at 14.5 GeV are consistent with the trend established by the results at lower and higher beam energies. The energy and centrality dependence of the baryon chemical potential ($\mu_{B}$), radial flow velocity ($\langle\beta\rangle$), and chemical and kinetic freeze-out temperatures are discussed systematically.
\end{abstract}
\end{frontmatter}

\section{Introduction}
\label{Introduction}
The main goal of the Beam Energy Scan (BES) program at RHIC is to study the structure of the Quantum Chromodynamics (QCD) phase diagram \cite{1}. The QCD phase diagram is usually plotted between the temperature and the baryon chemical potential ($\mu_{B}$). These thermodynamic quantities can be extracted from the measured hadron yields. The BES program has been carried out with the specific aim to explore several features of the QCD phase diagram such as to search for the phase boundary and the location of QCD critical point by colliding nuclei at several center-of-mass energies \cite{2}. In the years 2010 and 2011, data were collected for Au + Au collisions at $\sqrt{s_{NN}}$ = 7.7, 11.5, 19.6, 27, 39, 62.4 and 200 GeV. The corresponding $\mu_{B}$ ranged from 420 to 20 MeV, leaving a gap of about 100 MeV in the phase space between $\mu_{B}$ = 315 MeV and 205 MeV for $\sqrt{s_{NN}}$ = 11.5 and 19.6 GeV, respectively. This happened to be the region of phase space where several interesting observations related to bulk properties of the system were reported \citep{3,4,5}. In the year 2014, the data were recorded for Au+Au collisions at $\sqrt{s_{NN}}$ = 14.5 GeV (corresponding $\mu_{B}$ for central collisions $\approx$ 264 MeV) by the STAR detector. The significant hardware reconfiguration of the inner detectors of STAR are quite different from the 2010-2011 configuration. The fireball produced in heavy-ion collisions rapidly thermalizes leading to expansion and cooling of the system. During this process, two important stages occur as described below. The moment at which inelastic collisions cease is referred to as chemical freeze-out. The yields of most of the produced particles became constant at chemical freeze-out. The statistical thermal models have successfully described the chemical freeze-out stage with parameters such as chemical freeze-out temperature $T_{ch}$ and baryon chemical potential $\mu_{B}$. Even after the chemical freeze-out, the elastic interactions among the particles are still ongoing which could lead to changes in the momenta of the particles. After some time, when the inter-particle distance becomes so large that the elastic interactions stop, the system is said to have undergone kinetic freeze-out. At this stage, the transverse momentum ($p_{T}$) spectra of the produced particles became constant. The hydrodynamics inspired models such as the Blast Wave (BW) Model \citep{6,7,8} have described the kinetic freeze-out scenario with a common temperature $T_{kin}$ and average transverse radial flow velocity $\langle\beta\rangle$ which reflects the expansion in transverse direction.

Searching for the phase boundary in the QCD phase diagram is one of the main motivations of the BES program at RHIC. The elliptic flow ($v_{2}$) could be used as a powerful tool \cite{9}, because of the sensitivity of underlying dynamics in the early stage of the collisions. The Number of Constituent Quark (NCQ) scaling at the top RHIC energy ($\sqrt{s_{NN}}$ = 200 GeV) indicates that the collectivity has been built up at the partonic level \citep{10,11}. The BES results presented here are obtained for Au+Au collisions at $\sqrt{s_{NN}}$ = 14.5 GeV at mid-rapidity and compared with the results of other BES energies from $\sqrt{s_{NN}}$ = 7.7 - 39 GeV \citep{12,13}. The particle identification is done using both STAR Time Projection Chamber (TPC) \cite{14} and Time of Flight (TOF) \cite{15} detectors.

\section{Results and discussions}
\label{Results and discussions}
\subsection{Transverse momentum spectra}
\begin{figure}[h]
\centering
\includegraphics[totalheight=3.25cm]{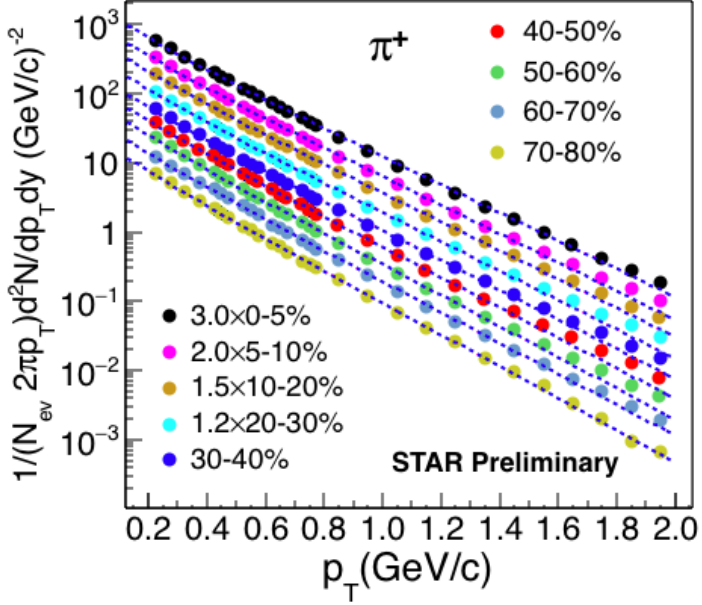}
\includegraphics[totalheight=3.25cm]{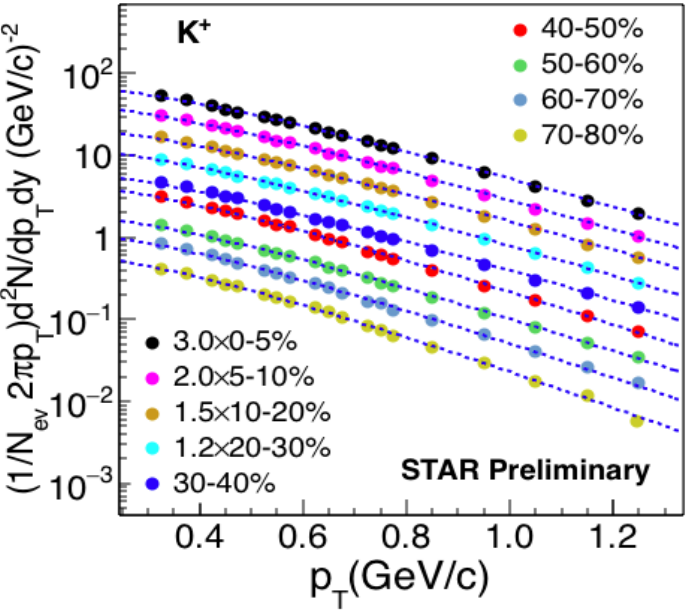}
\includegraphics[totalheight=3.25cm]{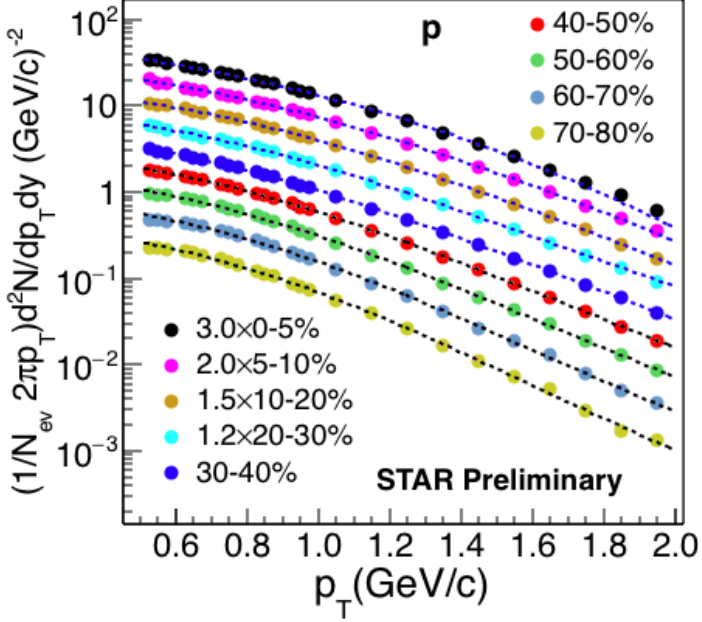}

\includegraphics[totalheight=3.3cm]{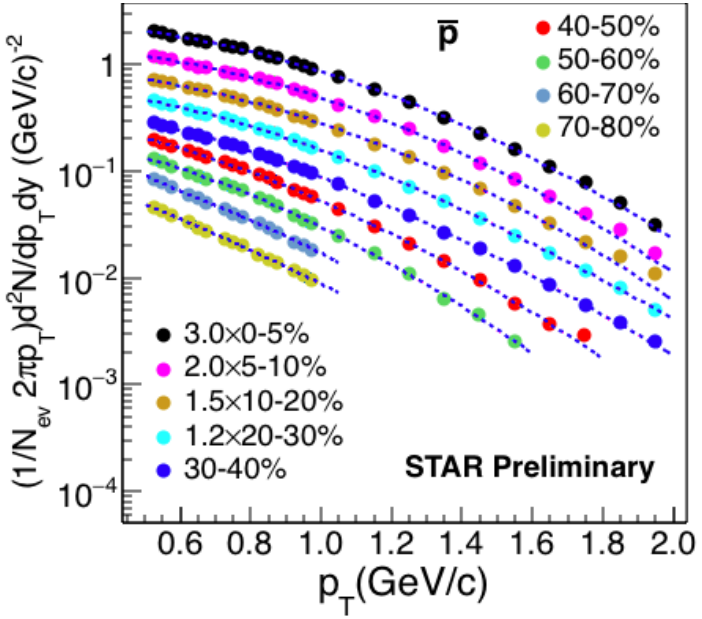}
\includegraphics[totalheight=3.3cm]{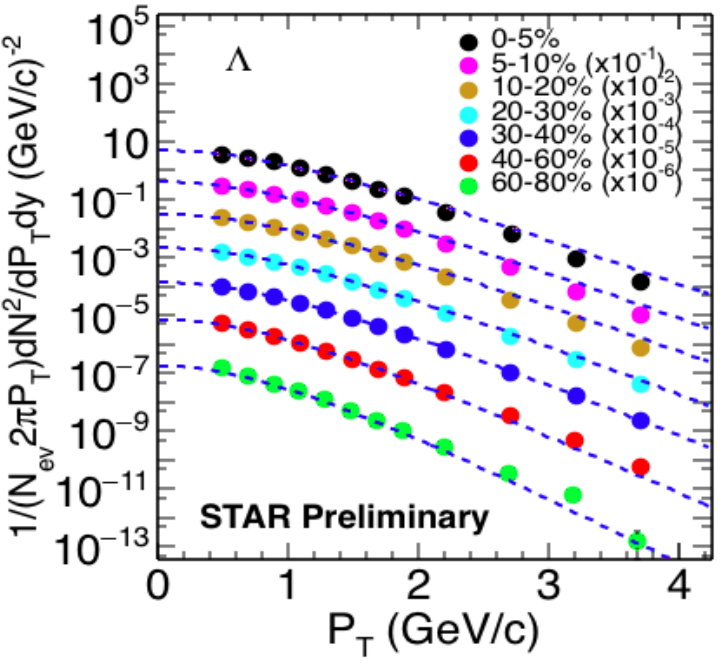}
\includegraphics[totalheight=3.3cm]{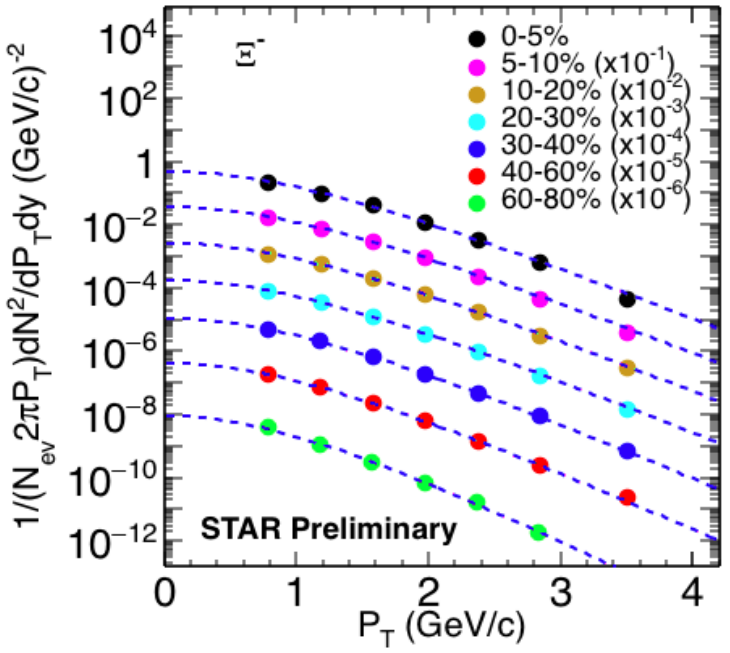}
\caption{\label{1} (Color online) $p_{T}$ spectra of $\pi^{+}$, $K^{+}$, $p(\bar{p})$, $\Lambda$ and $\Xi^{-}$ at mid-rapidity in Au+Au collisions at $\sqrt{s_{NN}}$ = 14.5 GeV. Errors shown are statistical only. The curves shown in each frame are for different centrality classes as labelled in the figure, and the different curves were scaled by factors also shown in the figure for clarity.} 
\end{figure}
Figures 1 shows the $p_{T}$ spectra of $\pi^{+}$, $K^{+}$, $p(\bar{p})$, $\Lambda$ and $\Xi^{-}$ in Au+Au collisions at $\sqrt{s_{NN}}$ = 14.5 GeV in different centralities at mid-rapidity. Errors shown are statistical only. The $p_{T}$-integrated yields are sum of two regions, measured region from data only, unmeasured region from fit extrapolation. The fit functions used to obtain yields are Bose-Einstein function ($A/(e^{m_{T}/T} -1)$) for  $\pi^{\pm}$, $m_{T}$-exponential function ($Ae^{-(m_{T}-m)/T}$) for  $K^{\pm}$, double exponential function ($A_{1}e^{-p_{T}^{2}/T_{1}^{2}} + A_{2}e^{-p_{T}^{2}/T_{2}^{2}}$) for $p(\bar{p})$ and Boltzmann function ($Am_{T}e^{-m_{T}/T}$) for $\Lambda$($\bar{\Lambda}$) and $\Xi^{\pm}$. The $\pi^{\pm}$, $K^{\pm}$, $p(\bar{p})$ yields are measured at rapidity $|y| < 0.1$ and those for $\Lambda$($\bar{\Lambda}$) and $\Xi^{\pm}$ are measured for $|y| < 0.5$. The $\pi^{\pm}$, $K^{\pm}$ and $p(\bar{p})$ yields are measured using only TPC for the $p_{T}$ range 0.2-0.8, 0.3-0.8 and 0.5-1.0 GeV/c, and using TPC + TOF for the $p_{T}$ range 0.4-2.0, 0.4-1.2 and 0.5-2.0 GeV/c, respectively. 

\subsection{Chemical freeze-out}
\begin{figure}[t]
\centering
\includegraphics[totalheight=3.35cm]{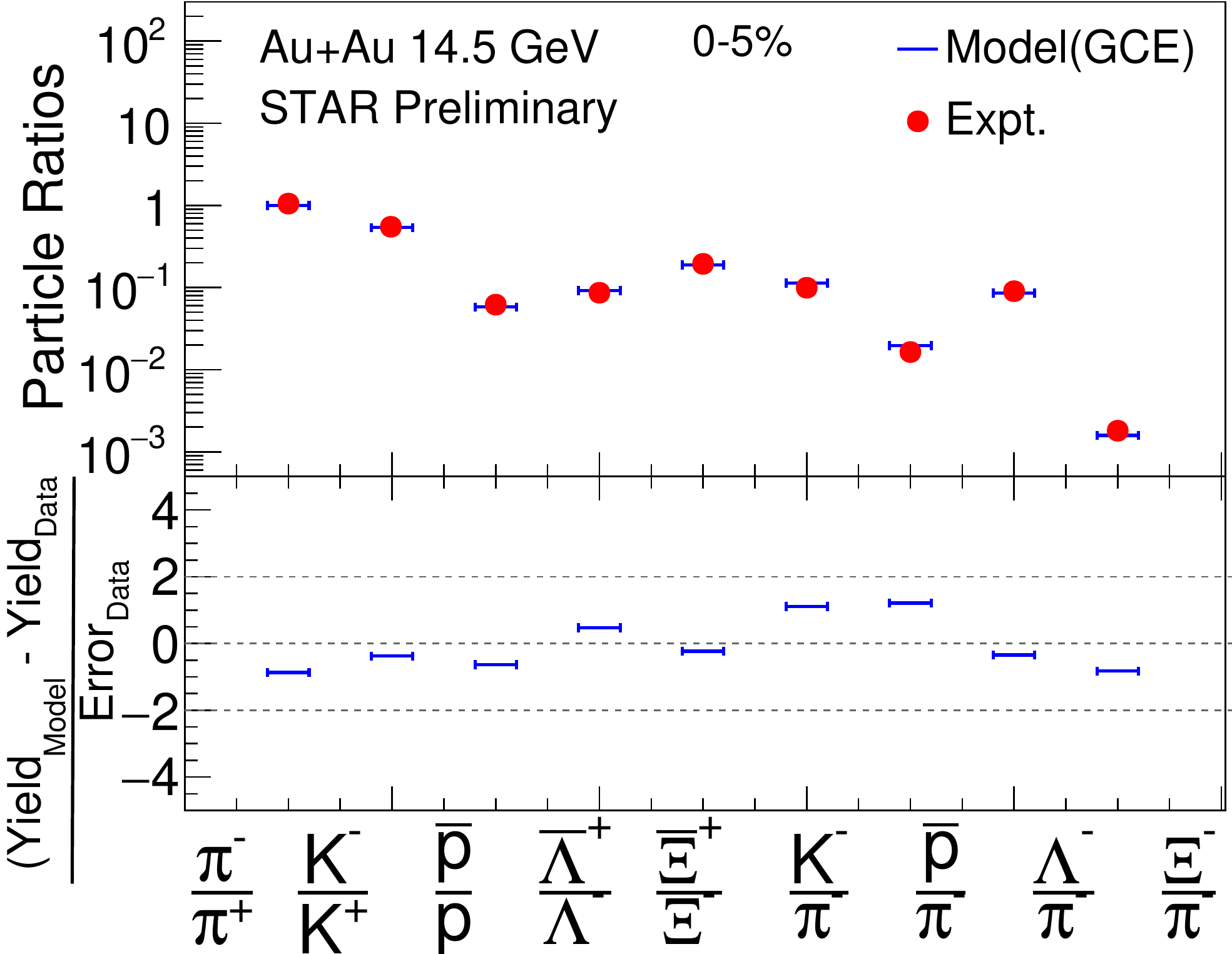} 
\hspace{0.5cm} \includegraphics[totalheight=3.3cm]{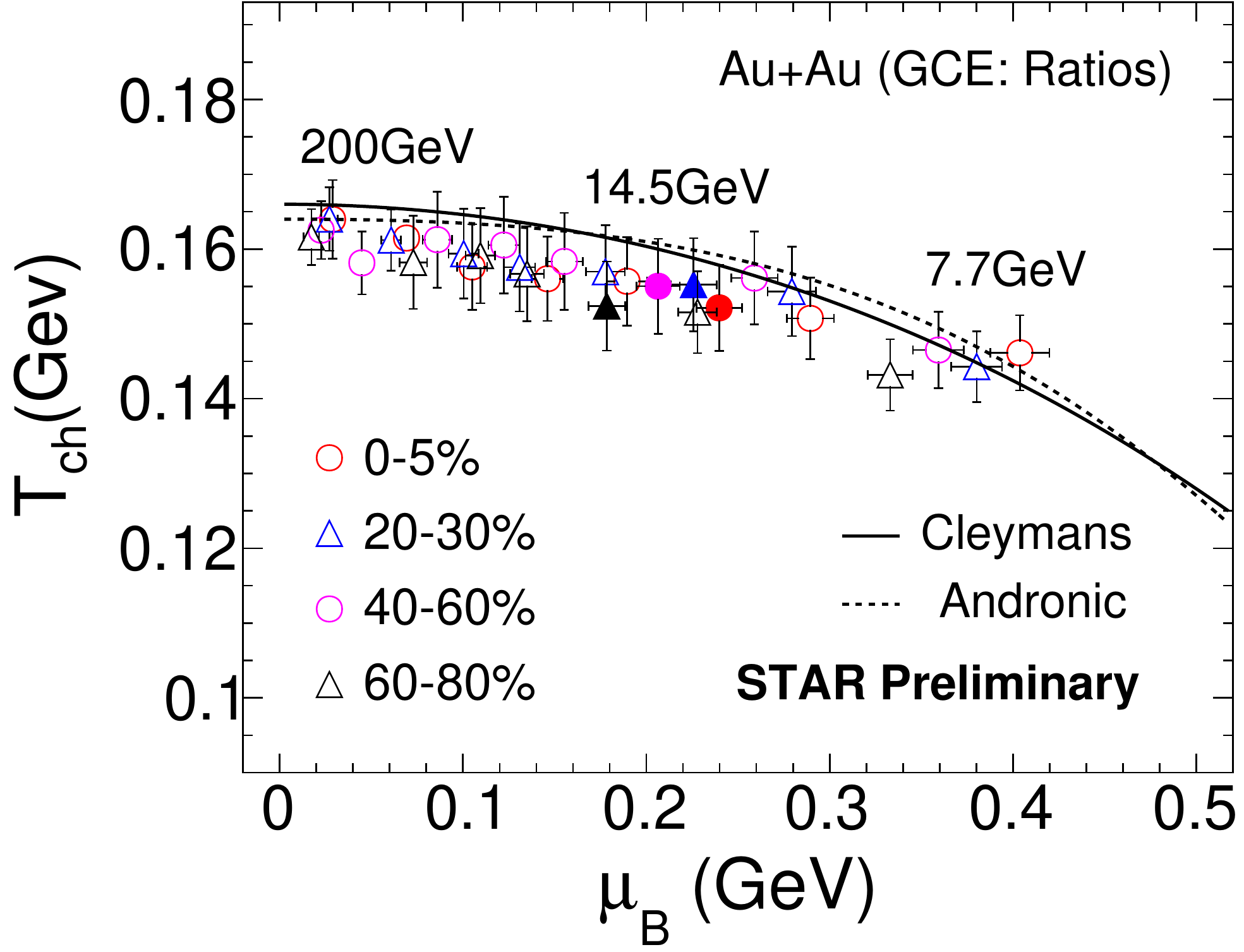}
\caption{\label{2} (Color online) (Left panel) Statistical thermal model fit to experimentally measured mid-rapidity particle ratios using grand-canonical ensemble for $0-5\%$ centrality in Au+Au collisions at $\sqrt{s_{NN}}$ = 14.5 GeV. (Right panel) $T_{ch}$ vs. $\mu_{B}$ from a statistical thermal model fit for Au+Au collisions at $\sqrt{s_{NN}}$ = 7.7$-$200 GeV.}
\end{figure}
Figure 2 (left panel) shows statistical thermal model fit to the particle ratios. Particle ratios are used in THERMUS model \cite{16} fit to get $T_{ch}$ and $\mu_{B}$ for different centralities in Au+Au collisions at $\sqrt{s_{NN}}$ = 14.5 GeV. Comparison of $T_{ch}$ and $\mu_{B}$ for different centralities with other BES energies is shown in the right panel of figure 2. The values of $T_{ch}$ and $\mu_{B}$ are consistent with the energy dependence trend of other energies. A centrality dependence of $\mu_{B}$ is observed, which is quite significant at lower energies. 

\subsection{Kinetic freeze-out}
\begin{figure}[h]
\centering
\includegraphics[totalheight=3.4cm]{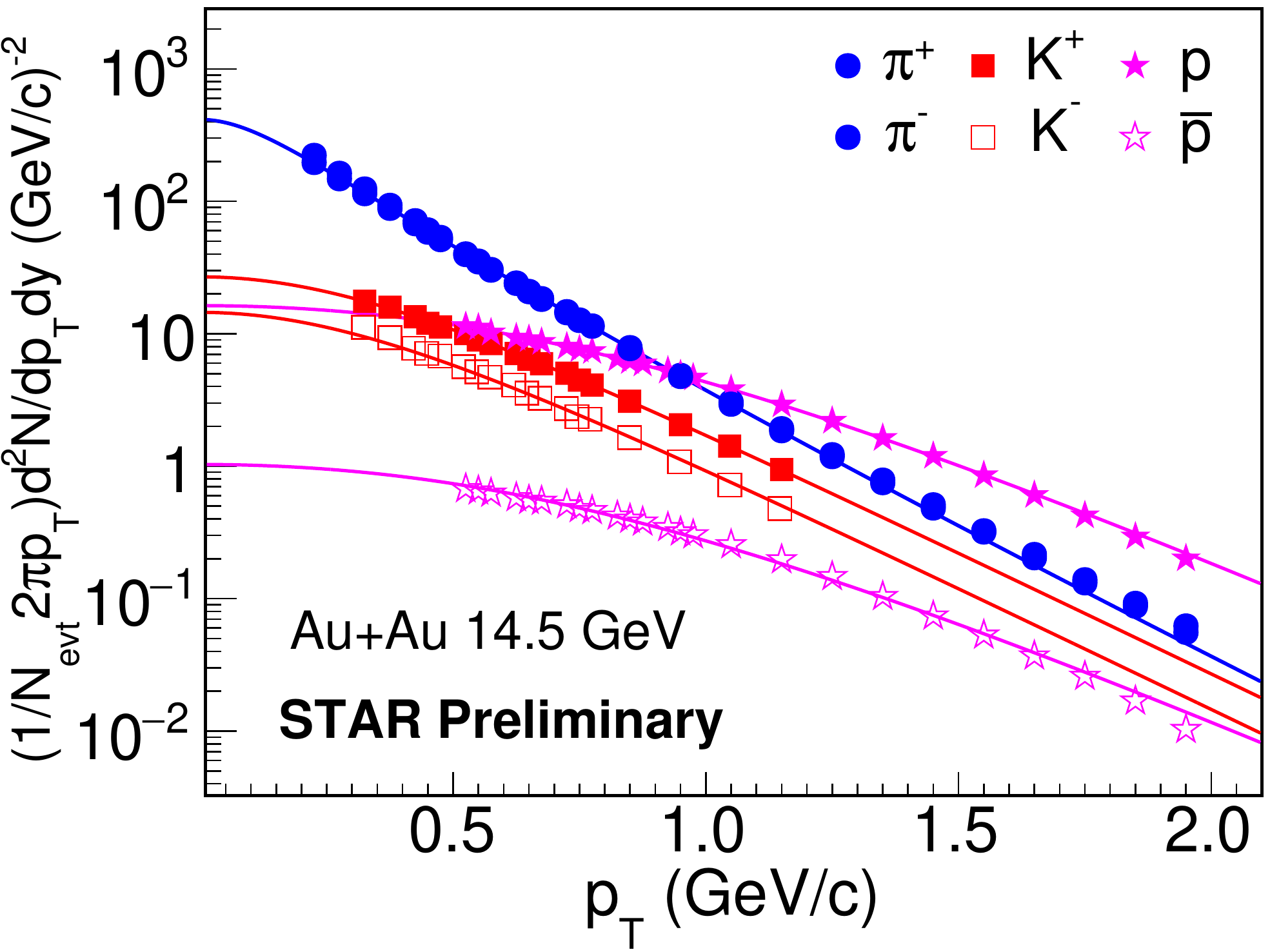} 
\hspace{0.5cm} \includegraphics[totalheight=3.4cm]{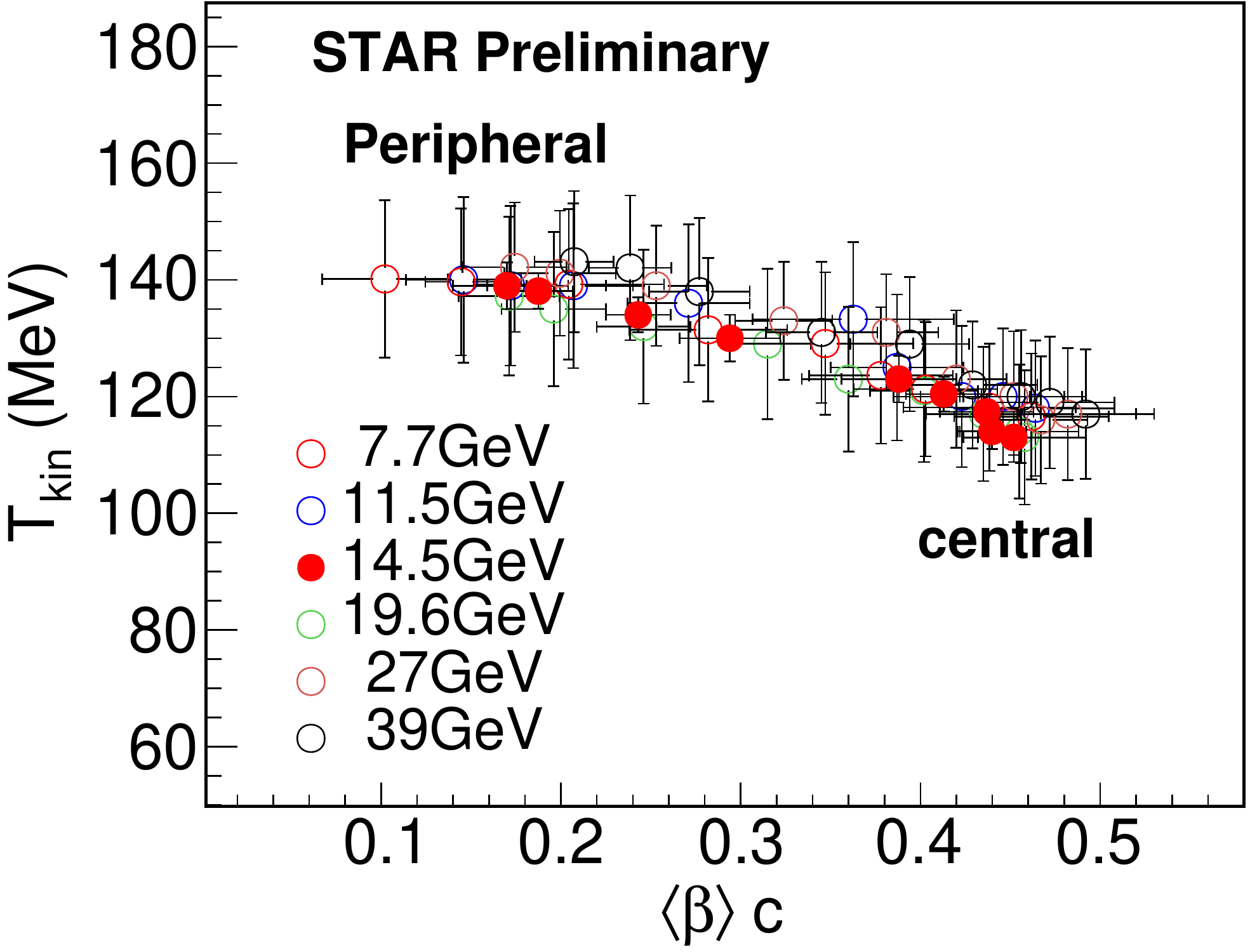}
\caption{\label{3} (Color online) (Left panel) Simultaneous BW fit to the $p_{T}$ spectra of $\pi^{\pm}$, $K^{\pm}$, $p(\bar{p})$ for $0-5\%$ centrality in Au+Au collisions at $\sqrt{s_{NN}}$ = 14.5 GeV. (Right panel) Variation of $T_{kin}$ with $<\beta>$ for different centralities at $\sqrt{s_{NN}}$ = 7.7$-$39 GeV. The errors shown here are statistical only.} 
\end{figure}
Kinetic freeze-out parameters are obtained using BW model \cite{4} by performing the simultaneous fits of $\pi^{\pm}$, $K^{\pm}$, $p(\bar{p})$, $p_{T}$ spectra as shown in the left panel of figure 3. The BW model describes the spectral shapes assuming a locally thermalized source with a common transverse flow velocity. The low $p_{T}$ part of pion spectra is affected by resonance decays due to which the pion spectra is fitted above $p_{T} > $ 0.5 GeV/c. Figure 3 (right panel) shows the variation of $T_{kin}$ as a function of $<\beta>$ for Au+Au collisions at  $\sqrt{s_{NN}}$ = 7.7$-$39 GeV. The $T_{kin}$ decreases from peripheral to central collisions. It also decreases with increasing collision energy. The $<\beta>$ increases with increase of energy as well as collision centrality. So, a higher value of $T_{kin}$ corresponds to lower value of $<\beta>$ and vice-versa.
\subsection{Elliptic flow ($v_{2}$)}
\begin{figure}[ht]
\centering
\includegraphics[totalheight=3.2cm]{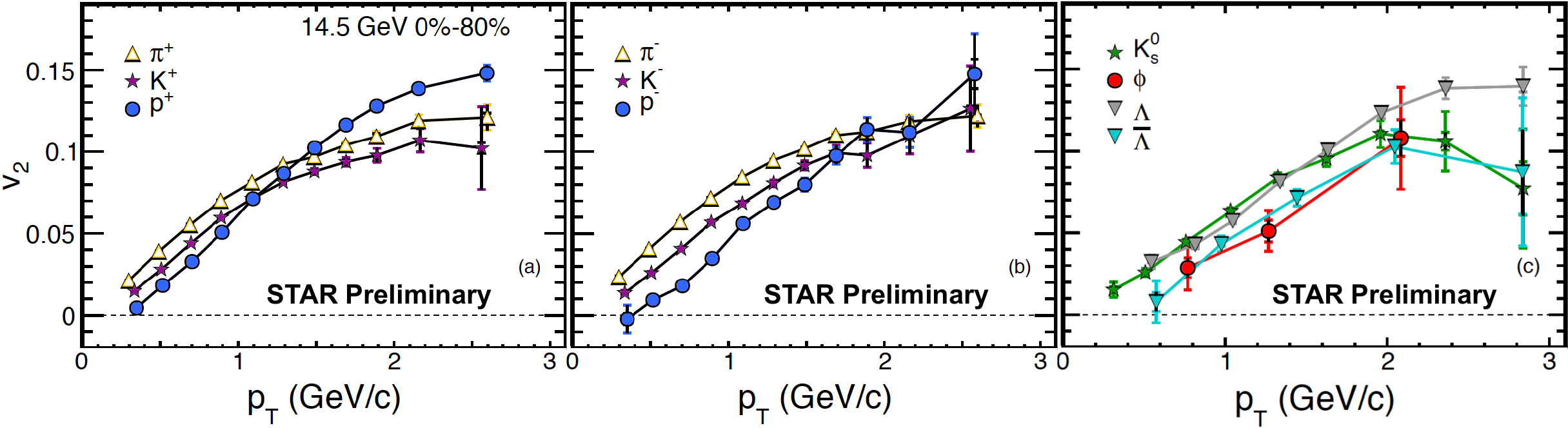} 
\caption{\label{4} (Color online) $v_{2}$ as a function of $p_{T}$ for the identified particles in minimum bias (0-80$\%$) Au+Au collisions at $\sqrt{s_{NN}}$ = 14.5 GeV. Error bars are shown for both the statistical and systematic uncertainties.} 
\end{figure}
Figure 4 shows the $v_{2}$ as a function of $p_{T}$ for the identified particles in minimum bias (0-80$\%$) Au+Au collisions at $\sqrt{s_{NN}}$ = 14.5 GeV. A clear mass ordering of $v_{2}$ at low $p_{T}$ ($<$ 1.5 GeV/c) is observed for $\pi^{+}$, $K^{+}$, p and their antiparticles while no distinct mass ordering is observed for $K_{s}^{0}$, $\phi$ and $\Lambda$($\bar{\Lambda}$). We have observed difference between the $v_{2}$ of $\Lambda$ and $\bar{\Lambda}$, which could be attributed to the difference in $v_{2}$ of particles and antiparticles \cite{12}. 
\begin{figure}[ht]
\centering
\includegraphics[totalheight=3.6cm]{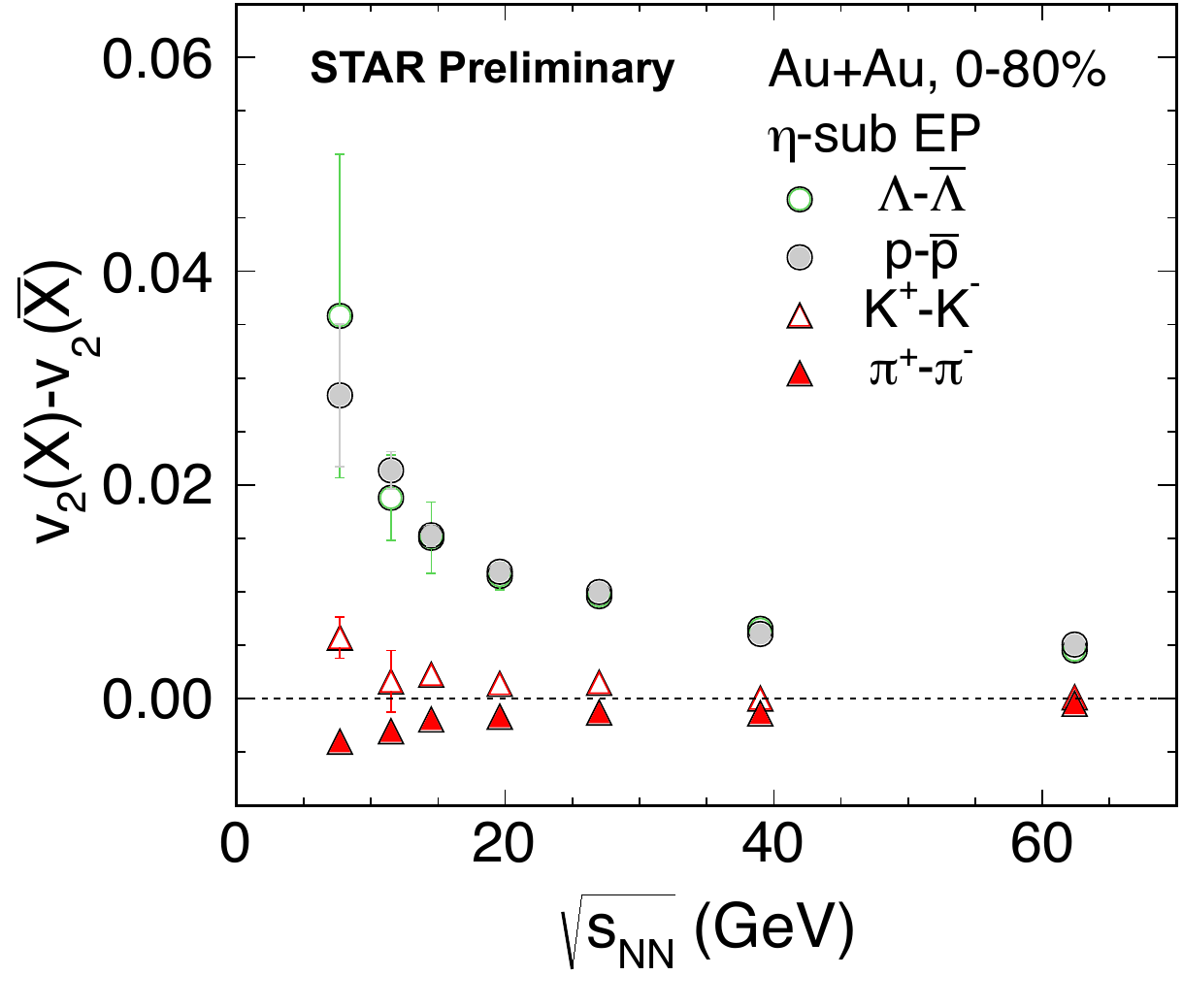} 
\caption{\label{5} (Color online) The difference in $v_{2}$ between particles and their corresponding anti-particles ($v_{2}(X)-v_{2}(\bar{X})$) as a function of beam energy in minimum bias ($0-80\%$) Au+Au collisions. The statistical uncertainties are shown by vertical lines.} 
\end{figure}
Figure 5 shows the difference in the average values of $v_{2}$ of particles and their corresponding anti-particles over measured $p_{T}$ range, defined as $\Delta v_{2} = v_{2}(X) - v_{2}(\bar{X})$. The data point for Au+Au collisions at $\sqrt{s_{NN}}$ = 14.5 GeV is preliminary measurement while the data points for other beam energies has been published \cite{17}. A monotonic increase of $\Delta v_{2}$ with decreasing collision energy is observed. The difference is more pronounced for baryons compared to mesons. The observed difference in $v_{2}$ reflects the breaking of NCQ scaling between particles and anti-particles, indicating that the contributions from hadronic interactions increase in the system evolution with decreasing collision energy \cite{17}.

\section{Summary}
\label{Summary}
We present $p_{T}$ spectra and $v_{2}$ of identified hadrons in Au+Au collisions at $\sqrt{s_{NN}}$ =14.5 GeV. The $T_{ch}$ versus $\mu_{B}$ shows the centrality dependence at lower energies. We observe that the chemical temperature $T_{ch}$ increases and the chemical potential $\mu_{B}$ decreases with increase in collision energy. We also observe centrality dependence of $T_{kin}$ and $\langle\beta\rangle$ and anti-correlation between them. Mass ordering for $v_{2}$ of $\pi^{+}$, $K^{+}$, p and their antiparticles is observed for low $p_{T}$ in Au+Au collisions at $\sqrt{s_{NN}}$ =14.5 GeV. The difference of $v_{2}$ between particles and corresponding anti-particles for $\pi$, $K$, $p(\bar{p})$ and $\Lambda(\bar{\Lambda})$ increases with decreasing beam energy. In summary, all the results of bulk property parameters and $v_{2}$ for Au+Au collisions at $\sqrt{s_{NN}}$ = 14.5 GeV are consistent with the trends established by the other BES energies.

\section{Acknowledgments}
\label{Acknowledgements}
We acknowledge the support by DAE-BRNS and DST/SERB Govt. of India.


\bibliographystyle{elsarticle-num}
\bibliography{<your-bib-database>}







\end{document}